\documentclass[12pt]{article}

\usepackage{amsmath,dsfont,natbib,graphicx}
\usepackage{epsfig,epstopdf}
\usepackage{graphicx}
\usepackage{psfrag}
\usepackage{amsmath}
\usepackage{amssymb,subfigure}
\usepackage{amsthm}
\usepackage{mathrsfs}
\usepackage{color}
\usepackage{algorithm2e}

\def\R{\mathbb R}

\def\F{\mathcal F}
\def\lp{\left(}
\def\rp{\right)}

\def\epsilon{\varepsilon}
\def\ds{\displaystyle}
\def\lp {\left( }
\def\rp {\right) }

\newcommand{\be}{\begin{equation}}
\newcommand{\ee}{\end{equation}}
\newcommand{\baa}{\begin{array}}
\newcommand{\eaa}{\end{array}}
\newcommand{\ba}{\begin{eqnarray}}
\newcommand{\ea}{\end{eqnarray}}
\newcommand{\bi}{\begin{itemize}}
\newcommand{\ei}{\end{itemize}}

\title{MULTILAND: a neutral landscape generator designed for theoretical studies}

\author{Lionel Roques  \\ \footnotesize{INRAE, BioSP, 84914, Avignon, France} \\ \footnotesize{lionel.roques@inrae.fr}}
\date{}

\begin{document}

\maketitle

\section{Presentation of the software}

The main goal of Multiland is to generate neutral landscapes made of several types of regions, with an exact control of the proportions occupied by each type of region \citep[see][for a review of current landscape modelling approaches]{GauHou14}. An important feature of the software is that it allows a control of the landscape fragmentation. It is intended to theoretical studies on the effect of landscape structure in applied sciences. 

\begin{figure}[h] \centering
\includegraphics*[width=0.6\textwidth]{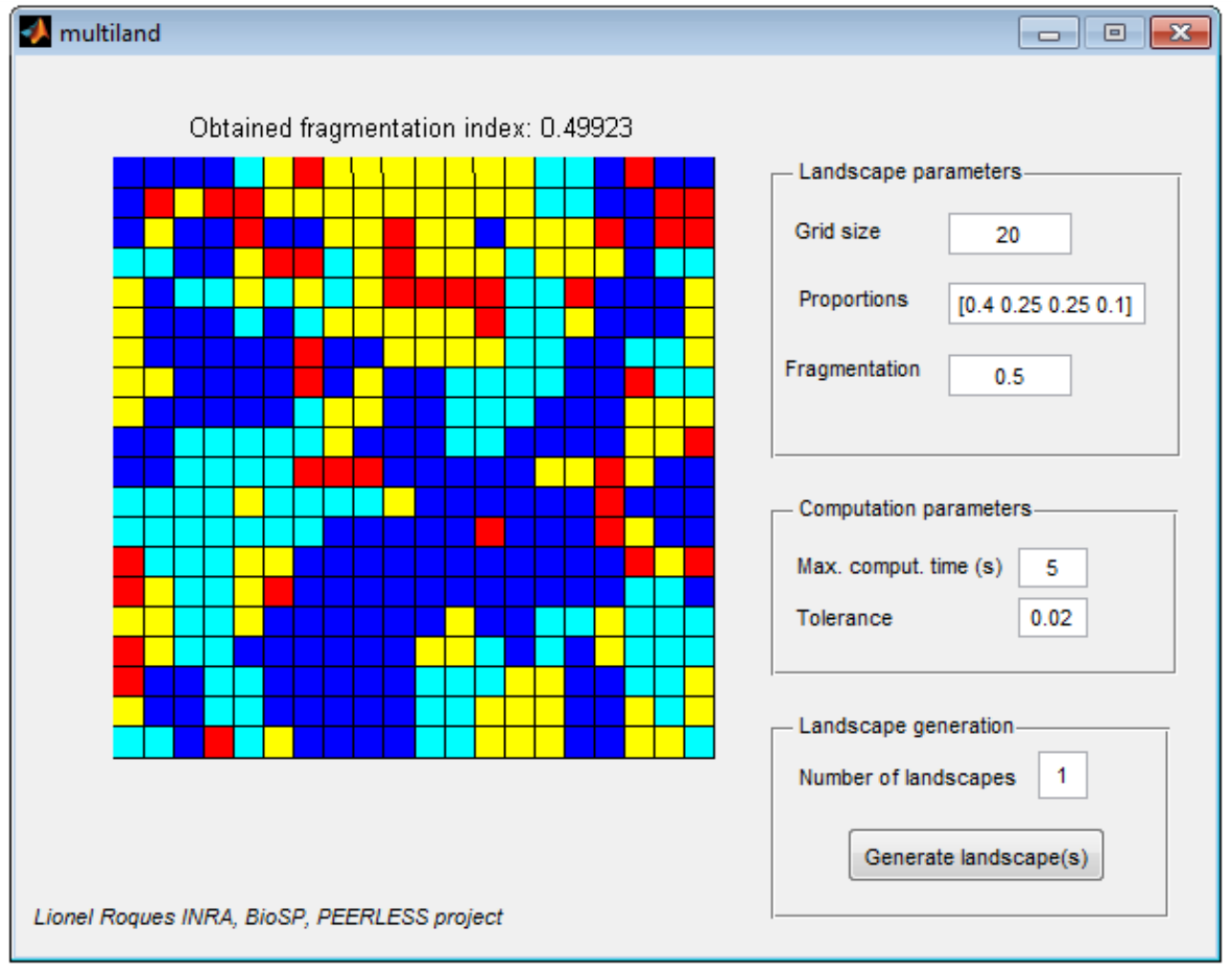}
\caption{User interface of the Multiland software.}%
\label{fig:soft} 
\end{figure}

\
There are two versions of the Multiland Software available: one for Matlab and one for Python (see Section~\ref{sec:down} for download details).
\begin{itemize}
    \item For Matlab, two options are provided. The first includes a user-friendly interface with 6 input fields (see Fig.~\ref{fig:soft}). The second is a version without a graphical interface, which may be more suitable for intensive computational tasks.
    \item For Python, the software is provided as a Jupyter notebook.
\end{itemize}

\clearpage

\noindent {\bf Inputs}

\

\noindent \textit{Landscape parameters:}
\bi
\item grid size: an integer $k\ge 2$. The landscape is described by a lattice of size $k\times k;$
\item proportions: a vector $V=(v_1, \ldots, v_M)$ of length $M\ge 2$. The length $M$ of the vector gives the number of different regions in the landscape.
The proportion of the lattice occupied by the region associated with a value $p\in [1,M]$ is equal to $\ds \mu_p= v_p/ \lp\sum\limits_{j=1}^{M}\rp$ whenever $\ds k^2\, \mu_p$ is an integer. Otherwise, the number of cells $n_p$ corresponding to region $p<M$ is given by the floor of $k^2 \, \mu_p:$ $\lfloor k^2 \, \mu_p \rfloor.$ The number of cells in the last region is given by $k^2- \sum\limits_{p=1}^{M-1}n_p.$
\item fragmentation index: a real in $[0,1].$ The value $0$ corresponds to the most aggregated landscapes; the value $1$ corresponds to the most fragmented landscapes.
\ei

\

\noindent \textit{Computation parameters:}
\bi
\item max. comput. time: number of seconds allowed for the generation of $1$ landscape. Default value is $5$s. Computation time depends on the grid size; on the value of the fragmentation index (longer times are necessary for values close to $0$ or $1$); and on the tolerance (see below);
\item tolerance: the computation is stopped when the maximum computation time is reached or when the  algorithm has generated a landscape with the desired fragmentation index $\pm$ the tolerance.
\ei

\

\noindent \textit{Landscape generation:}
\bi
\item Number of landscapes: once the button ``Generate landscape(s)" is pressed, the indicated number of landscapes is generated. The matrixes describing the landscapes are save in a file ``U.txt". The corresponding fragmentation indexes are save in a file ``fr.txt".
\ei

\section{Theoretical aspects}

The software Multiland is based on a stochastic landscape model with $M$ types of regions. This model should:
\bi
\item[-] allow an exact control of the proportion attributed to each region;
\item[-] lead to more or less fragmented landscapes, with a control of the degree of fragmentation.
\ei

\

\noindent {\bf State space.} Let $\Xi$ be the lattice made of the of
the cells $(i,j)$, $i\in \{1,\ldots,k\}$, $i\in \{1,\ldots,k\}$, equipped with a neighboring system $V_\Xi$. We
consider the random field $\omega$ defined on the lattice $\Xi$.
This random field assigns each site $\xi=(i,j)$ ($\xi\in\Xi$)
a value $p$ in $\{1,2,\ldots,M\}$, while the number
$n_{p}=\hbox{card}\{\omega_\xi=p, \ \xi \in \Xi\}$, takes a
pre-fixed value. Under these assumptions let $\Omega$ be the
corresponding state space.

\

\begin{figure} \centering
\includegraphics*[width=0.7\textwidth]{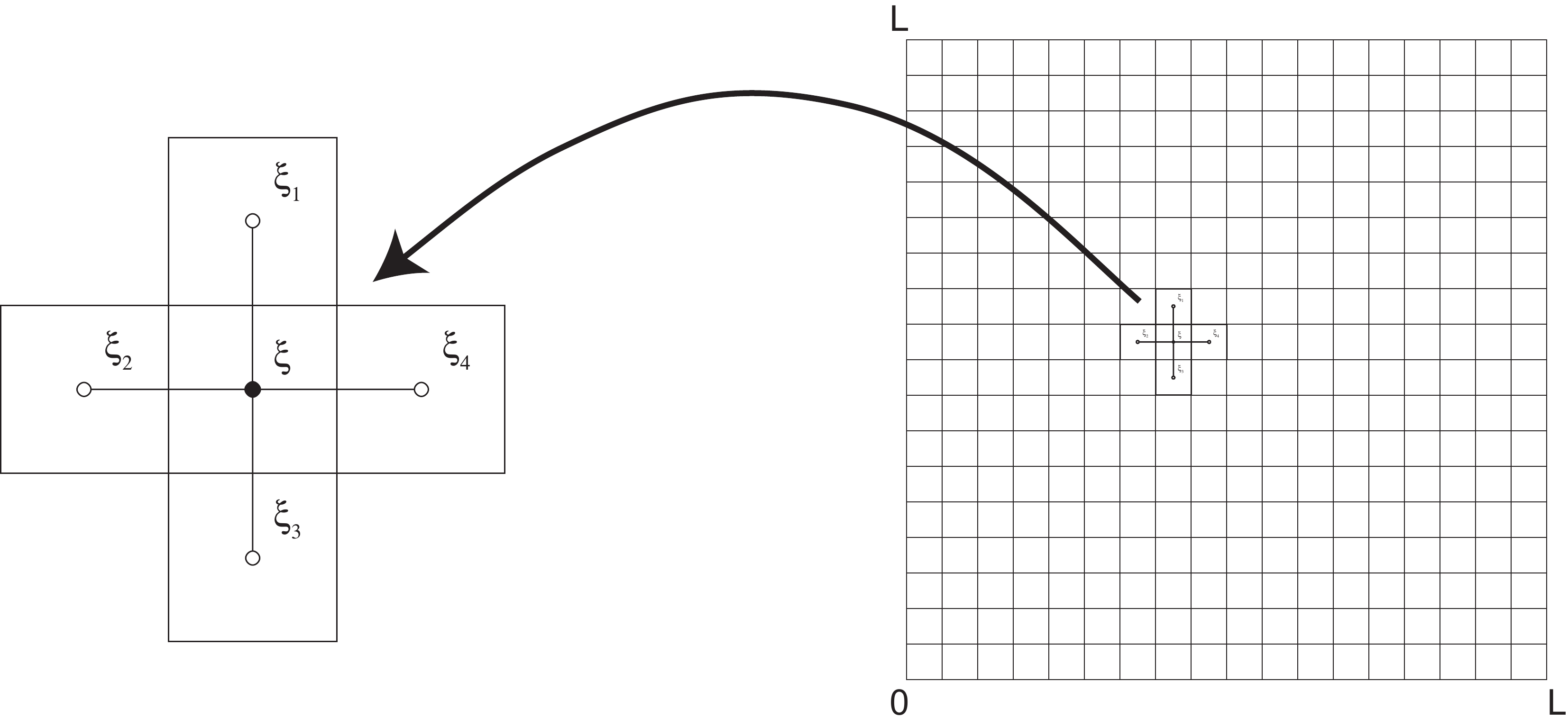}
\caption{The 4-neighbourhood system: an element $\xi$ of $\Xi$ and its four neighbours $\xi_i$, $i=1\ldots 4$.}%
\label{fig:vois4} 
\end{figure}

\noindent {\bf Neighborhood system.}  We adopt  a 4-neighborhood system as shown in Fig.~\ref{fig:vois4}. In order to take account of the periodicity of
the environment, the domain is considered wrapped on a torus.

\

\noindent {\bf Fragmentation index.}
 For each $\omega$ in $\Omega$, we set
$$\ds{s(\omega)=\frac{1}{2}\sum_{\eta, \xi \in \Xi, \ \eta \in
V_\xi}
\mathds{1}\{\omega_\xi=\omega_\eta\}},$$ the number of pairs of
 neighbors $\xi,\eta\in\Xi$ such that $\omega_\xi=\omega_\eta$ ($\mathds{1}\{\cdot\}$
 is the indicator function). The statistic of the model, $s(\omega)$, is
 directly linked to the habitat fragmentation: a landscape pattern
 is all the more aggregated as $s(\omega)$ is high, and all the
 more fragmented as $s(\omega)$ is small. This statistic will be used below in the construction of the landscape model.

In order to obtain a fragmentation index which belongs to $[0,1]$ ($0$: most aggregated landscapes, $1$: most fragmented landscapes), we proceed as follows. First, we choose $q$ in $\{1,2,\ldots,M\}$ such that $n_q\ge n_p$ for all $p\in \{1,2,\ldots,n\}.$ The region associated with the value $q$ corresponds to the ``background" of our landscape. Then, for each other region $z$ associated with a value in $I=\{1,\ldots,q-1,q+1, \ldots n\}$, we define the fragmentation level
\be
f_z=1-\frac{s_z(\omega)}{B(n_z)},
\ee
where $\ds{s_z(\omega)=\frac{1}{2}\sum_{\eta, \xi \in \Xi, \ \eta \in
V_\xi}
\mathds{1}\{\omega_\xi=\omega_\eta=z\}},$ corresponds to the number of pairs of similar neighbors in the region $z$ and $B(n_z)=2\, n_z -\hbox{ceil} (2\, \sqrt{n_z})$ corresponds to the maximum possible number of similar neighbors in a region made of $n_z$ cells. Thus, $f_z\in [0,1].$ Finally, we set
\be \label{eq:frag_index}
\F(\omega)=\frac{1}{n-1}\sum_{z\in I}{f_z} \in [0,1],
\ee
and we say that $\F(\omega)$ is the \textit{fragmentation index} of the landscape $\omega.$

\

\noindent {\bf Gibbs measure.} The landscape model is based on the Gibbs measure $P$, defined over
$\Omega$ by \be P(X=\omega)=\frac{1}{Z}e^{-U(\omega)},\label{eq:prob1}\ee with the
partition function $\ds{ Z= \sum_{\omega \in
\Omega}e^{-U(\omega)}}$ and the Gibbs energy function $U: \Omega \to
\R$,
 \be \label{eq:mod1}
U(\omega)=-\beta s(\omega), \ee with $\beta \in \R$;  thus, the
 probability density becomes \be \label{eq:pmod1} P(X=\omega)=\frac{1}{Z}e^{\beta
 s(\omega)}.\ee The proposed landscape model favours the pairs of
 neighbours with the same value, while $\beta$ increases. Hence,
 $\beta$ directly controls the topology of the landscape patterns.

\

\noindent {\bf Generation of the landscapes.} Samples from the distribution \eqref{eq:pmod1} are obtained by using a Metropolis-Hastings algorithm. More precisely,
starting from a random initial landscape, a Metropolis-Hastings algorithm is run until (1) the maximum allowed computation time is reached or (2) the distance between the fragmentation index of the current state and the desired fragmentation index is smaller than the tolerance (see Section~1). The parameter $\beta$ is adjusted during the algorithm in order to increase or decrease the fragmentation index, depending on the desired fragmentation index.

\section{References and contact}
A simpler version of the algorithm was used in \cite{RoqSto07} to generate binary landscapes. This new version has been developed in the framework of the PEERLESS ANR project ``Predictive Ecological Engineering for Landscape Ecosystem Services and Sustainability". For more information, contact the author at lionel.roques@inrae.fr.

\section{Downloading the software \label{sec:down}}
The Matlab$^\circledR$ source code is available at http://multiland.biosp.org.

\end{document}